\documentclass[12pt,twoside,a4paper]{article}
\usepackage{amssymb,amsmath}
\usepackage{amsfonts}
\usepackage{indentfirst}
\usepackage{graphicx}
\newcommand{\ys}{\overline{y}}
\newcommand{\zs}{\overline{z}}
\begin{document}
{\Large\bf Riccati equation\,-\,based generalization of  \par
Dawson's integral function} \vspace{0.5cm}
\begin{quote}{\small\begin{center}R. Messina, M.A. Jivulescu, A. Messina and A. Napoli\end{center}
 \textit{MIUR, CNISM and Dipartimento di Scienze Fisiche ed
Astronomiche,\\ Universit\`{a} di Palermo, via Archirafi 36, 90123
Palermo, Italy}}\end{quote}

\begin{center}
\textmd{SUMMARY}\end{center}A new generalization of Dawson's
integral function based on the link between a Riccati nonlinear
differential equation and a second-order ordinary differential
equation is reported. The MacLaurin expansion of this generalized
function is built up and to this end an explicit formula for a
generic cofactor of a triangular matrix is deduced.

\begin{center}\textmd{1. INTRODUCTION}\end{center}
\par The study of homogeneous linear second-order ordinary
differential equations (HLIIODE) is strictly related to that of
Riccati equations. It is possible to express the general integral
of an arbitrary HLIIODE in terms of a particular solution of the
associated Riccati equation. In the first part of this paper we
recall this procedure in order to write down explicitly the most
general solution of a HLIIODE vanishing at a conveniently chosen
fixed point. The structure of such a general solution suggests a
very simple way to extend Dawson's integral function (DIF),
appearing in various physical contexts such as spectroscopy,
electrical oscillations, heat conduction, astrophysics and so on
as well as in applied mathematics \cite{Garcia}-\cite{Schreier}.
Quite recently the DIF has emerged in the treatment of the
dynamics of the so\,-\,called generalized spin star system where
the reduced dynamics of a system composed of two central qubits is
investigated in the limit of an infinite number of environmental
spins \cite{Petruccione}. The DIF deserves, in addition, attention
on the mathematical side since it is related to other special
functions and even because it possesses useful formal properties
exploitable for its numerical evaluations \cite{Cody,McCabe}. The
main scope of this paper is to propose a new class of
transcendental functions which may be viewed as generalizations of
the DIF. We report in detail the construction of the MacLaurin
expansion of a generic element of this class by solving a linear
system of infinitely many equations in infinitely many unknowns.
We reach this goal evaluating all the cofactors of a triangular
matrix of arbitrary finite dimension having all its diagonal
elements equal to one.

\begin{center}\textmd{2. RICCATI GENERAL SOLUTION OF A
HLIIODE}\end{center}\vspace{0.5cm}
\par Consider the following homogenous second-order linear
differential equation with variable coefficients
\begin{equation}\label{EqPrinc}y''(x)+b_1(x)y'(x)+b_2(x)y(x)=0\end{equation}
where $b_1(x)$ and $b_2(x)$ are $C^2$ functions on
$(-a,a)\in\mathbb{R}$, $a>0$. Such a linear equation is invariant
under dilatations $y\mapsto\lambda y$ the infinitesimal generator
of which is
\begin{equation}X=y\,\frac{\partial}{\partial y}.\end{equation}
Lie theory of symmetry of differential equations takes into account
such symmetry and, as indicated in \cite{Carinena}, the recipe is to
look for a new coordinate $u$ such that $X$ becomes
\begin{equation}X=\frac{\partial}{\partial u}.\end{equation}
The coordinate $X$ is such that $Xu=1$, i.e. $y\frac{\partial
u}{\partial y}=1$, from which we obtain $y(x)=e^{u(x)}$ (up to
multiplication by a constant).

Since in terms of the new coordinate we get
\begin{equation}y'(x)=u'(x)y(x),\qquad y''(x)=\Bigl[u''(x)+\Bigl(u'(x)\Bigr)^2\Bigr]y(x)\end{equation}
the transformed differential equation may be cast in the following
form:
\begin{equation}\label{Riccatiu}u''(x)+\Bigl(u'(x)\Bigr)^2+b_1(x)u'(x)+b_2(x)=0\end{equation}
which takes the form of a Riccati equation
\begin{equation}\label{Riccatiz}z'(x)+z^2(x)+b_1(x)z(x)+b_2(x)=0\end{equation}
if we put
\begin{equation}\label{zu}z(x)=u'(x).\end{equation}

Let $\zs(x)$ be a particular solution of eq.\eqref{Riccatiz}. Then
on integrating eq.\eqref{zu} one immediately gets the particular
solution of eq.\eqref{EqPrinc}
\begin{equation}y_1(x)=e^{\int_0^x\zs(t)dt}\end{equation}
such that $y_1(0)=1$ and $y_1'(0)=\zs(0)$.

We recall that, once a solution is known, we can introduce the
change of variable $y(x)=y_1(x)v(x)$ and then, taking into account
that
\begin{equation}y'(x)=y'_1(x)v(x)+y_1(x)v'(x)\qquad y''(x)=y''_1(x)v(x)+2y_1'(x)v'(x)+y_1(x)v''(x)\end{equation}
and the fact that $y_1(x)$ is a solution of eq.\eqref{EqPrinc},
eq.\eqref{EqPrinc} becomes
\begin{equation}\label{Equazv}v''(x)=a(x)v'(x)\end{equation}
where we have defined
\begin{equation}\label{Posa}a(x)=-\Bigl(2\zs(x)+b_1(x)\Bigr).\end{equation}
A solution of eq.\eqref{Equazv} is
\begin{equation}v(x)=\int_0^x\exp\Biggl(\int_0^ta(t')dt'\Biggr)dt\end{equation}
from which we obtain the following solution of eq.\eqref{EqPrinc}:
\begin{equation}y_2(x)=y_1(x)\int_0^x\exp\Biggl[\int_0^ta(t')dt'\Biggr]dt\end{equation}
satisfying the initial conditions $y_2(0)=0$ and $y_2'(0)=1$.

\noindent In view of the fact that
\begin{equation}W(0)\equiv\left(
\begin{array}{ccccccc}
y_1(0)&y_2(0)\\
y_1'(0)&y_2'(0)\\
\end{array}\right)=
\left(\begin{array}{cc}
1&0\\
\zs(0)&1\\
\end{array}\right)=1\neq0,
\end{equation}
we immediately write down the following general solution of
eq.\eqref{EqPrinc}:
\begin{equation}\label{SolGen}y(x)=\exp\left[\int_0^x\zs(t)dt\right]\left[C_1+C_2\int_0^x\exp\left[-\int_0^t\Bigl(2\zs(t')+b_1(t')\Bigr)dt'\right]dt\right].\end{equation}
Requiring $y(0)=0$ yields $C_1=0$ so that the most general solution
$\ys(x)$ of eq.\eqref{EqPrinc} vanishing at $x=0$ may be cast in the
form
\begin{equation}\label{ysegn}\ys(x)=C\exp\left[\int_0^x\zs(t)dt\right]\cdot\int_0^x\exp\left[-\int_0^t\Bigl(2\zs(t')+b_1(t')\Bigr)dt'\right]dt\end{equation}
which satisfies the further Cauchy condition $\ys'(0)=C$.
\vspace{1cm}
\begin{center}\textmd{3. EXTENDING DAWSON'S INTEGRAL FUNCTION}\end{center}
\vspace{0.5cm}
\par The form of $\ys(x)$ provides a very favorable starting
point to construct generalizations of Dawson's integral odd
function
\begin{equation}\label{Dawor}F(2,x)=\exp(-x^2)\int_0^x\exp(t^2)dt=-F(2,-x)\end{equation}
where $x\in\mathbb{R}$. In the literature there have appeared
extensions like
\begin{equation}\label{Dawp}F(p,x)=\exp(-x^p)\int_0^x\exp(t^p)dt\end{equation}
with $p=2,3,4,\dots$ and $x\in\mathbb{R}$, in connections with
interesting problems both in applied Physics and in Mathematics
\cite{Sajo,Dijkstra}. Our scope is to exploit the structure of
$\ys(x)$ to propose new generalizations of the original Dawson's
integral function. The first natural extension suggested by
equations\eqref{Dawor} and \eqref{Dawp} is to look for $b_1(x)$ in
equation (11) such that
\begin{equation}\label{Posiz}\zs(x)=2\zs(x)+b_1(x)\quad\Rightarrow\quad\zs(x)=-b_1(x)\end{equation}
This condition establishes the following link between $b_1(x)$ and
$b_2(x)$ throughout the associated Riccati equation \eqref{Riccatiz}
\begin{equation}-b_1'(x)=-b_1^2(x)+b_1^2(x)-b_2(x)\Rightarrow b_2(x)=b_1'(x)\end{equation}
so that we may claim that the unique solution of the Cauchy problem
\begin{equation}\left\{\begin{array}{l}
y''(x)+b(x)y'(x)+b'(x)y(x)=y''(x)+\Bigl(b(x)y(x)\Bigr)'=0\\
y(0)=0\qquad y'(0)=1\\\end{array}\right.\end{equation} may be
written down as
\begin{equation}\label{Sol12}y(x)=\exp\bigl[-B(x)\bigr]\int_0^x\exp\bigl[B(t)\bigr]dt\end{equation}
where
\begin{equation}B(x)=\int_0^xb(t)dt.\end{equation}
When $b(x)=2x$ or $b(x)=px^{p-1}$ eq.\eqref{Sol12} gives back the
function $F(2,x)$ and $F(p,x)$ respectively. We denote the
function defined by eq.\eqref{Sol12}, by $D_b(x)$ and call it the
generalized Dawson's integral function associated to the function
$b(x)$.

There is another way of extending Dawson's integral function
easily suggested by  equation \eqref{ysegn}. If we in fact
stipulate that
\begin{equation}\left\{\begin{array}{ll}
\zs(x)=-\lambda px^{p-1} & p\in\mathbb{N}^+,\lambda\in\mathbb{R}-\{0\}\\
2\zs(x)+b_1(x)=-\mu sx^{s-1}\qquad &
s\in\mathbb{N}^+,\mu\in\mathbb{R}-\{0\}\\\end{array}\right.\end{equation}
then $b_1(x)=2\lambda px^{p-1}-\mu sx^{s-1}$, and consequently
\begin{equation}b_2(x)=\lambda^2p^2x^{2p-2}-\lambda\mu psx^{p+s-2}+\lambda
p(p-1)x^{p-2}.\end{equation}Thus we may claim that the function
\begin{equation}F\bigl[(\lambda,p),(\mu,s);x\bigr]\equiv\exp\bigl(-\lambda x^p\bigr)
\int_0^x\exp\bigl(\mu t^s\bigr)dt\end{equation} with
$p,s\in\mathbb{N}^+$, $\lambda,\mu\in\mathbb{R}-\{0\}$,
$x\in\mathbb{R}$ is the unique solution of the following Cauchy
problem
\begin{equation}\left\{\begin{array}{l}
y''(x)+\Bigl(2\lambda px^{p-1}-\mu
sx^{s-1}\Bigr)y'(x)+\\
\qquad+\Bigl(\lambda^2p^2x^{2p-2}-\lambda\mu
spx^{s+p-2}+\lambda p(p-1)x^{p-2}\Bigr)y(x)=0\\
y(0)=0\qquad y'(0)=1\\\end{array}\right. .\end{equation}
$F[(1,p),(1,p);x]$ is of course coincident with $F(p,x)$ and is
included in eq.\eqref{Sol12} in correspondence with
$b(x)=px^{p-1}$. \vspace{1cm}
\begin{center}\textmd{4. MACLAURIN EXPANSION OF $D_b(x)$}\end{center}
\vspace{0.5cm}
\par We concentrate on $D_b(x)$ posing the following
question: if $b(x)$ may be expanded in a MacLaurin series with the
 radius of convergence $R$, that is
\begin{equation}b(x)=\sum_{n=0}^{+\infty}\frac{b^{(n)}(0)}{n!}x^n,\end{equation}
$b^{(n)}(x)$ being the $n$-th derivative of $b(x)$, is it possible
to find the MacLaurin expansion of $D_b(x)$ in terms of the class
of coefficients $\bigl\{b^{(n)}(0),n\in\mathbb{N}\bigr\}$\,? The
question is well posed, since the assumption on $b(x)$ guarantees
that $D_b(x)$ too may be expanded in MacLaurin series with the
same  radius of convergence $R$. Then our problem is to build up
the explicit expression of $D_b^{(k)}(0)$ as a function of
$\bigl\{b^{(n)}(0),n\in\mathbb{N}\bigr\}$. To this end we begin by
observing that for any $b(x)$
\begin{equation}D_b'(x)=-D_b(x)b(x)+1,\qquad D_b''(x)=-\bigl[D_b(x)b(x)\bigr]'\end{equation}
\begin{equation}D_b(0)=0,\qquad D_b'(0)=1\end{equation}
\begin{equation}D_b^{(k+2)}(x)=-\bigl[D_b(x)b(x)\bigr]^{(k+1)}=-\sum_{n=0}^{k+1}{k+1\choose n}D_b^{(n)}(x)b^{(k+1-n)}(x)\end{equation}
\begin{equation}\label{Sistema}D_b^{(k+1)}(0)=-kb^{(k-1)}(0)-\sum_{n=2}^{k}{k\choose n}D_b^{(n)}(0)b^{(k-n)}(0).\end{equation}

\noindent To find $D_b^{(k+1)}(0)$, with $k\in\mathbb{N}^+$, we must
solve the linear system of $k$ equations in the $k$ unknowns
$\Bigl\{D_b^{(2)}(0),D_b^{(3)}(0),\dots,D_b^{(k+1)}(0)\Bigr\}$. Its
incomplete matrix is
\begin{equation}A_k=\left(\begin{array}{ccccccc}
1&&&&&&\\
{2\choose2}b(0)&1&&&&&\\
{3\choose2}b^{(1)}(0)&{3\choose3}b(0)&1&&&&\\
{4\choose2}b^{(2)}(0)&{4\choose3}b^{(1)}(0)&{4\choose4}b(0)&1&&&\\
{5\choose2}b^{(3)}(0)&{5\choose3}b^{(2)}(0)&{5\choose4}b^{(1)}(0)&{5\choose5}b(0)&1&&\\
\dots&\dots&\dots&\dots&\dots&\ddots&\\
{k\choose2}b^{(k-2)}(0)&{k\choose3}b^{(k-3)}(0)&{k\choose4}b^{(k-4)}(0)&\dots&\dots&{k\choose
k}b(0)&1\\\end{array}\right)
\end{equation}
and has lower triangular form with all its diagonal elements equal
to one.

\noindent The vector $B_k$ of the constants appearing in the linear
system \eqref{Sistema} is
\begin{equation}B_k=\left(\begin{array}{c}
-b(0)\\
-{2\choose1}b^{(1)}(0)\\
-{3\choose1}b^{(2)}(0)\\
\dots\\
-{k\choose1}b^{(k-1)}(0)\\
\end{array}\right).
\end{equation}

\noindent It is convenient to introduce the following notation for
the elements of $a_{ij}$ of $A_k$ and $b_i$ of $B_k$:
\begin{equation}a_{i,j}=\left\{
\begin{array}{lll}
0&&i<j\\
1&&i=j\\
{i\choose j+1}b^{(i-j-1)}(0)&&i>j\\
\end{array}\right.\end{equation}
\begin{equation}b_i=-{i\choose 1}b^{(i-1)}(0)=-ib^{(i-1)}(0)\end{equation}
provided that $1\leq i,j\leq k$. Since $\det A_k=1$ for any
$k\in\mathbb{N}^+$, Cramer's theorem yields
\begin{equation}D_b^{(k+1)}(0)=\det P_k\end{equation}
where
\begin{equation}P_k=\left(\begin{array}{cccccc}
1&0&0&0&\dots&-b(0)\\
{2\choose2}b(0)&1&0&0&\dots&-2b^{(1)}(0)\\
{3\choose2}b^{(1)}(0)&{3\choose3}b(0)&1&0&\dots&-3b^{(2)}(0)\\
\dots&\dots&\dots&\dots&\vdots\\
{k\choose2}b^{(k-2)}(0)&{k\choose3}b^{(k-3)}(0)&\dots&\dots&{k\choose k}b(0)&-kb^{(k-1)}(0)\\
\end{array}\right).
\end{equation}
The determinant associated to the matrix $P_k$ borders the
determinant of $A_{k-1}$ through the addition of the $k$-th row and
the $k$-th column. The value of $\det P_k$ may thus be evaluated by
means of the well known Cauchy formula \cite{VeinDale} according to
which
\begin{equation}\det B=\det\left(\begin{array}{ccccc}
c_{11}&c_{12}&\dots&c_{1n}&\alpha_1\\
c_{21}&\dots&\dots&\dots&\alpha_2\\
\dots&\dots&\dots&\dots&\vdots\\
c_{n1}&\dots&\dots&c_{nn}&\alpha_n\\
\beta_1&\beta_2&\dots&\beta_n&a\\
\end{array}\right)=a\det C-\sum_{r,s=1}^n\alpha_r\beta_s\Delta_{r,s}\end{equation}
where $\Delta_{r,s}$ is the cofactor of the corresponding element
$c_{r,s}$ of the bordered matrix
\begin{equation}C=\left(\begin{array}{cccc}
c_{11}&c_{12}&\dots&c_{1n}\\
c_{21}&\dots&\dots&\dots\\
\dots&\dots&\dots&\dots\\
c_{n1}&\dots&\dots&c_{nn}\\
\end{array}\right).\end{equation}
In our case $B=P_k$ and $C=A_{k-1}$, so that this formula assumes
the form
\begin{eqnarray}D^{(k+1)}(0)=\det P_k=-kb^{(k-1)}(0)+\sum_{i,j=1}^{k-1}i\binom{k}{j+1}b^{(k-j-1)}(0)b^{(i-1)}(0)\Delta_{i,j}\end{eqnarray}
$\Delta_{i,j}$ being the cofactor of the correspondent element
$a_{i,j}$ ($1\leq i,j\leq k-1$) of the matrix $A_{k-1}$. It is
easy to convince oneself that $\Delta_{i,j}=0$ for any $i>j$,
since the elimination of the $i$-th row and $j$-th column with
$i>j$ gives a triangular matrix with one diagonal element equal to
$0$. It is in addition evident that $\Delta_{i,i}=1$ for any $i$.
Thus we have only to evaluate the cofactors
$C(i,n)\equiv\Delta_{i,i+n}$ with $1\leq i\leq k-2$ and $1\leq
n\leq k-i-1$ finding the explicit expression of such cofactors for
a generic lower triangular matrix $C=(c_{r,s})$ of order $k-1$
with all its diagonal elements equal to $1$. We indeed claim that
\begin{equation}\label{Cofatt}C(i,n)=\sum_{s=1}^n(-1)^s\sum_{\{p\}_s^n}c_{i+n,i+p_1}c_{i+p_1,i+p_2}\dots
c_{i+p_{s-2},i+p_{s-1}}c_{i+p_{s-1},i+p_s}\end{equation} where
$\sum_{\{p\}_s^n}$ denotes $\sum_{n>p_1>p_2>\dots>p_{s-1}>p_s=0}$
that is summation over all the $\binom{n-1}{n-s}$ possible sets of
$s$ indices $\{p_1,p_2,\dots,p_s\}$, taken in strict decreasing
order and such that $0<p_i<n$, $i=1,2,\dots,s-1$ and $p_s=0$. We
observe that the terms of eq.\eqref{Cofatt} corresponding to a fixed
value of $s$ are products of $s$ elements of the matrix $A$ with a
plus (minus) sign if $s$ is even (odd). Then, the only term
corresponding to $s=1$ is $-c_{i+n,i}$.

Let us prove eq.\eqref{Cofatt} by induction on $n$ for an
arbitrary $i=1,2,\dots,k-2$. Firstly we show that
\begin{equation}C(i,1)=-c_{i+1,i}.\end{equation}
To obtain $C(i,1)=\Delta_{i,i+1}$ we have to calculate and multiply
by $-1$ the determinant of the matrix obtained from $A_{k-1}$
eliminating the $i$-th row and the $(i+1)$-th column. It is easy to
observe that this operation leaves a triangular matrix having
$c_{i+1,i}$ as a diagonal element, and all other diagonal elements
equal to $1$. As a consequence, we have $\Delta_{i,i+1}=-c_{i+1,i}$,
as stated before. Let us now suppose that, for all $m\leq n$ we have
\begin{equation}\label{IpInd}C(i,m)=\sum_{s=1}^m(-1)^s\sum_{\{p\}_s^m}c_{i+m,i+p_1}c_{i+p_1,i+p_2}\dots
c_{i+p_{s-2},i+p_{s-1}}c_{i+p_{s-1},i+p_s}.\end{equation} Using
 Laplace's second theorem with the $i$-th row for the cofactors and
the $(i+n+1)$-th row for the elements we obtain
\begin{equation}\label{Laplace}\sum_{j=1}^{k-1}c_{i+n+1,j}\Delta_{i,j}=0.\end{equation}
Using the fact that $C(i,0)=1$, $c_{i,j}=0$ for $j>i$ whereas
$c_{i,i}=1$ we obtain
\begin{equation}c_{i+n+1,i}+\sum_{j=i+1}^{i+n}c_{i+n+1,j}C(i,j-i)+C(i,n+1)=0.\end{equation}
Using eq.\eqref{IpInd} and posing $j-i=m$ we have
\begin{eqnarray}C(i,n+1)=-c_{i+n+1,i}+\sum_{m=1}^n\sum_{s=1}^m(-1)^{s+1}\sum_{\{p\}_s^m}c_{i+n+1,i+m}c_{i+m,i+p_1}\dots
c_{i+p_{s-1},i+p_s}.
\end{eqnarray}
Observing that for a fixed value of $s$ the second term contains
products of $s+1$ elements $c_{i,j}$ with the correct sign factor,
it is easy to convince oneself that we may write
\begin{eqnarray}C(i,n+1)=-c_{i+n+1,i}+\sum_{s=2}^{n+1}(-1)^s\sum_{\{p\}_s^{(n+1)}}c_{i+n+1,i+p_1}c_{i+p_1,i+p_2}\dots
c_{i+p_{s-1},i+p_s}\end{eqnarray} that is
\begin{equation}C(i,n+1)=\sum_{s=1}^{n+1}(-1)^s\sum_{\{p\}_s^{(n+1)}}c_{i+n+1,i+p_1}c_{i+p_1,i+p_2}\dots
c_{i+p_{s-1},i+p_s}\end{equation} which concludes the demonstration
eq.\eqref{Cofatt}.

As far as we know the explicit expression of all the cofactors of a
triangular determinant of arbitrary finite order having all its
diagonal elements equal to one has not previously appeared in
literature.

\noindent Summing up we have proved that, for any $1\leq i,j\leq
k-1$,
\begin{equation}\Delta_{i,j}=\left\{\begin{array}{ll}
C(i,j-i)&i<j\\
1&i=j\\
0&i>j\\
\end{array}\right..\end{equation}
This concludes the derivation of the MacLaurin expansion of
$D_b(x)$ for any $b(x)$ as well expandable in MacLaurin series. We
guess that other general properties may be proved for such class
of extended Dawson's integral functions also thanks to the
explicit knowledge of their MacLaurin expansions. Moreover special
cases obtained in correspondence to particular choices of the
function $b(x)$ may find applications both in Physics and in
Applied Mathematics.

\begin{center}\textmd{5. CONCLUDING REMARKS}\end{center}
\par The structure of the general integral of a linear homogeneous
second-order ordinary differential equation in the form given by
eq.\eqref{SolGen} suggests a very simple way to generalize
Dawson's integral function. Such a generalization introduces
indeed a class of new functions all possessing the same
Dawson\,-\,like structure.

In order to explore some properties of this class without choosing
the form of $b(x)$, we have coped with the interesting question of
finding the MacLaurin expansion of $D_b(x)$ in terms of the
MacLaurin coefficients of the same $b(x)$. It is worth noting that
the resolution of this problem has lead us to derive the explicit
expression of all the cofactors of a lower triangular matrix which
diagonal entries are all equal. We underline that such a
derivation may be successfully generalized and applied to a
generic triangular matrix. In conclusion we feel that the ideas
and methods reported in this paper may be of some help to propose
new generalizations of other special functions of interest in
physics, chemistry and applied mathematics.


\begin{thebibliography}{7}{\small{
\bibitem{Garcia} Garcia T.T. Voigt profile fitting to quasar absorption lines: an analytic approximation to the Voigt-Hjerting function \emph{Monthly Notices of the Royal Astronomical Society} \textbf{369},
2025-2035, 2006
\bibitem{Casini} Casini R. The Hanle Effect of the Two-Level Atom in the Weak-Field Approximation  \emph{The  Astrophysical Journal} \textbf{568}, 1056-1065, 2002
\bibitem{Kaiser}  Kaiser A. et al. Microscopic processes in dielectrics under irradiation by subpicosecond laser pulses
\emph{Physical Review B} \textbf{61}, 11437-11450, 2000
\bibitem{Wei} Peng-Sheng Wei et al. Distribution functions
of positive ions and electrons in a plasma near a surface
\emph{IEEE Transactions on Plasma Science} \textbf{28}, 1244-1253,
2000
\bibitem{Stolzt}  Cody W.J. and  Stoltz  I. The use of Taylor series to test accuracy of function programs
 \emph{ACM Transactions on Mathematical Software (TOMS)}
\textbf{17}, 55 - 63, 1991
\bibitem{DiRocco}  Di Rocco H.O. and  Aguirre Tellez M. Evaluation of the asymmetric Voigt profile and complex error functions in terms of the Kummer functions
\emph{Acta Physica Polonica A} \textbf{106}, 817-827, 2004
\bibitem{Shippony}  Shippony Z. and  Read W.G. A very accurate algorithm for the Voigt profile \emph{Journal of
Quantitative Spectroscopy and Radiative Transfer} \textbf{50},
635-546, 1993
\bibitem{Wendlandt} Wendlandt B.C.H. Temperature in an irradiated thermally conducting translucent medium \emph{Journal of Physics D: Applied
Physics} \textbf{6}, 657-660, 1973
\bibitem{Lehle}  Lehle H. and al. Probing electric fields in protein cavities by using the vibrational Stark effect of carbon monoxide  \emph{Biophysical Journal}  \textbf{88}, 1978-1990, 2005
\bibitem{Schreier}  Schreier F. Voigt and complex error function:
a comparison of computational methods \emph{Journal of
Quantitative Spectroscopy and Radiative Transfer} \textbf{48},
743-762, 1992
\bibitem{Petruccione} Hamdouni Y.,  Fannes M. and  Petruccione F.  Exact dynamics of a two-qubit system in a spin star environment  \emph{Physical Review B }\textbf{73}, 245323-12, 2006
\bibitem{Cody}
 Cody W.J.,  Paciorek K.A. and  Thacher H.C. Chebyshev approximations for Dawson's integral \emph{Mathematics of Computation} \textbf{24}, 171-178, 1970
\bibitem{McCabe}  McCabe J.H. A continued fraction expansion, with a truncation error estimate, for Dawson's integral
 \emph{Mathematics of Computation} \textbf{28}, 811-816, 1974
\bibitem{Carinena}Cari\~{n}ena J.F. and  Ramos A. A new geometric approach to Lie
Systems and Physical Applications \emph{Acta Applicandae
Mathematicae} \textbf{70}, 43-69, 2002
\bibitem{Sajo}
 Sajo E. On the recursive properties of Dowson's integral \emph{Journal of Physics A: Mathematical and General}  \textbf{26}, 2977-2987, 1993
\bibitem{Dijkstra}
 Dijkstra D. A continued fraction expansion for a generalization of Dawson's integral \emph{Mathematics of Computation} \textbf{31}, 503-510,
 1977
\bibitem{VeinDale}
 Vein R. and  Dale P.  \emph{Determinants and their Applications in
Mathemathical Physics}, Springer, New York, 1999}}
\end{thebibliography}
\end{document}